# Results of Complex Observations of Asteroid (596) Scheila at the Sanglokh International Astronomical Observatory

G. I. Kokhirova[a, *], O. V. Ivanova[b, c], F. Dzh. Rakhmatullaeva[a], U. Kh. Khamroev[a], M. Buriev[a], and S. Kh. Abdulloev[a]

[a]*Institute of Astrophysics, Academy of Sciences, Republic of Tajikistan, Dushanbe, 734042 Republic of Tajikistan*
[b]*Astronomical Institute, Slovak Academy of Sciences, Tatranská Lomnica, 05960 Slovak Republic*
[c]*Main Astronomical Observatory, National Academy of Sciences of Ukraine, Kyiv, 03143 Ukraine*
*\*e-mail: kokhirova2004@mail.ru*

**Abstract**—Results of astrometric and BVRI photometric observations of the active asteroid (596) Scheila are presented. The observations were carried out at the Zeiss-1000 telescope of the Sanglokh International Astro-nomical Observatory of the Institute of Astrophysics of the Academy of Sciences of the Republic of Tajikistan on June 16–17 and from July 30 to August 1, 2017. The coordinates of the object and its orbit were determined; and the apparent brightness in four filters, the absolute brightness in the $V$ and $R$ filters, and the color indices were obtained. The light curves suggest that no substantial changes in the asteroid's brightness occurred during the observations. The absolute brightness of the asteroid in the $V$ and $R$ filters was $(9.1 \pm 0.05)^m$ and $(8.8 \pm 0.03)^m$, respectively. The mean value of the asteroid diameter was $(119 \pm 2)$ km. The mean values of the color indices ($B-V = (0.72 \pm 0.05)^m$, $V-R = (0.29 \pm 0.03)^m$, and $R-I = (0.31 \pm 0.03)^m$) agree well with the values for asteroids of the P- and D-types and its averages. The rotation period of the asteroid estimated from photometric observations was $16.1 \pm 0.2$ h. The analysis of the data has shown that the asteroid contin-ues to exhibit the same values of absolute brightness and other characteristics as those before the collision with a small body in December 2010, though the latter resulted in the outburst event and cometary activity of the asteroid. Most likely, the collision of asteroid (596) Scheila with a small body did not lead to catastrophic changes in the surface of the asteroid or to its compete break-up.

## INTRODUCTION

The population of small bodies of the Solar System includes comets, asteroids, and meteoroids. In recent years, the objects of a transitional class, which is between comets and asteroids, were found in the main asteroid belt (MAB). Objects of this group exhibit the dynamical characteristics that are typical of asteroids of the main belt: their orbits are within Jupiter's orbit, while Tisserand's parameter $T_j > 3$; at the same time, they show the signs of atmospheres and activity, which is inherent in comets. This recognized class of objects is called main-belt comets (MBCs) or active asteroids (Jewitt et al., 2009; Hsieh et al., 2009a, 2009b). There is a small fraction of so-called "dormant" or "extinct" comets among the near-Earth asteroids, and the problems connected with their study have existed for a long time (Opik, 1963; Babadzhanov and Kokhirova, 2009). First of all, the study of these objects provides one of few opportunities to determine directly the physical properties of cometary nuclei. The other objects of the Solar System that probably combine the properties of both atmosphereless bodies and comets are centaurs. However, the dynamical characteristics of extinct comets and centaurs substantially differ from those of the transitional objects of the main belt. The detection of cometary activity in asteroid (596) Scheila and other objects confirms the existence of the comet–asteroid transition class (Larson, 2010; Larson et al., 2010; Jewitt, 2012; Jewitt et al., 2016). To date, the following comets of the MAB and active asteroids are known from observations: (3200) Phaethon, 311P/PanSTARRS (P/2013 P5), P/2010 A2 (LINEAR), (1) Ceres, (2201) Oljato, P/2012 F5 (Gibbs), 259P/Garradd (P/2008 R1), (596) Scheila, (62412) 2000 SY178, P/2013 R3 (Catalina-PanSTARRS), 133P/(7968) Elst–Pizarro, 176P/LINEAR(118401), 238P/Read (P/2005 U1), P/2012 T1 (PanSTARRS), 313P/Gibbs (P/2014 S4), 324 P/La Sagra (P/2010 R2), and 107P/(4015) Wilson–Harrington (Jewitt, 2012; Jewitt et al., 2016). The existence of the MBCs is important, because they apparently form the third res-ervoir of comets in the Solar System (after the Oort cloud and the Kuiper belt) (Jewitt et al., 2009; Hsieh et al., 2009a, 2009b). The comparison of the objects from three reservoirs allows the protoplanetary disk of the Sun to be studied in three regions: comets (asteroids) of the main belt, which are at a distance of approximately 2.2–3.6 AU, where they were formed under a temperature of ~150–200 K depending on the distance to the Sun and the surface properties (Masiero et al., 2012); comets of the Kuiper belt formed at distances of 30 to 50 AU under an equilibrium temperature of 30–40 K; and the Oort cloud of comets formed at distances of 50 to 100 AU under an equilibrium temperature of approximately 10 K (Jewitt, 2015). Thus, to determine the optical and physical characteristics of the MBCs, to reveal the real nature of bursts of cometary activity in asteroids, which manifests itself in ejecting dust and forming typical cometary tails, and to find their relation to the other bodies of the Solar System is currently an urgent task. Such studies are linked to the fundamental problem of the origin and interconnection of small bodies in the Solar System.

**Table 1.** The main parameters of asteroid (596) Scheila*

| Orbital characteristics (J2000.0) | | Physical characteristics | |
|---|---|---|---|
| Epoch | Sept. 22, 2017 | Absolute brightness $H$ | $8.9^m$ |
| $a$ | 2.927 AU | Apparent brightness $m$ | $(11.7-15.3)^m$ |
| $q$ | 2.449 AU | Diameter $D$ | $(159.7 \pm 1.1)$ km (Masiero et al., 2012) |
| $Q$ | 3.406 AU | Rotation period $n$ | 15.848 h |
| $e$ | 0.163 | Geometric albedo $p_v$ | $0.038 \pm 0.004$ (Tedesco, Desert, 2002) |
| $i$ | 14.661° | Color index $B-V$ | $0.714^m$ |
| $\omega$ | 175.157° | Color index $U-B$ | $0.177^m$ |
| $\Omega$ | 70.606° | Orbital period, $P$ | 5.01 yr |
| $T_j$ | 3.209 | | |

* All of the parameters in Table 1, except those provided with the references, are given according to the NASA database (http://ssd.jpl.nasa.gov; 2017).

## ACTIVE ASTEROID (596) SCHEILA

The active asteroid (596) Scheila (afterwards, Scheila) was found in photographic plates by A. Kopff from Heidelberg on February 21, 1906. The main orbital and physical characteristics of the asteroid are listed in Table 1: the semimajor axis $a$, the perihelion $q$ and aphelion $Q$ distances, the eccentricity $e$, the inclination $i$, the argument of perihelion $\omega$, the longitude of ascending node $\Omega$, Tisserand's parameter $T_j$, the value of which indicates an orbit typical of asteroids, and the orbital period $P$.

On December 11, 2010, during observations at the Mount Lemmon Observatory carried out within the Catalina Sky Survey frames (Arizona, United States), it was found that the brightness of asteroid Scheila increased almost twofold, a cometary-like coma was formed, and then three dust tails appeared (Larson, 2010; Larson et al., 2010). The image of asteroid (596) Scheila obtained at the Catalina Sky Survey's 152.4-cm telescope during the outburst in December 2010 is shown in Fig. 1 (the authors are A. Gibbs and S. Larson; credit https://uanews.arizona.edu). The image was obtained by combining 30 exposures of the asteroid.

Since the object exhibited the signs of cometary activity, it was classified as an active asteroid. The detected cometary activity, which is not inherent in asteroids, attracted much attention to the object. The asteroid was studied on the basis of the data of the UV–Optical Telescope onboard the *Swift* spacecraft (Bodewits et al., 2011) and the Hubble Space Telescope (Jewitt et al., 2011) and from the ground-based observations (see, e.g., Betzler et al., 2012; Kiselev et al., 2014; Neslushan et al., 2016). A substantial reddening in the UV range was found in the spectrum of the asteroid. The measurements showed that the asteroid's outburst was sporadic and unrelated to the gas release typical of comets (Bodewits et al., 2011). It was shown that the asteroid ejected $6 \times 10^8$ kg of dust with a velocity of 57 m/s. The most probable cause of the abrupt activity of asteroid Scheila is a collision with an unknown MAB asteroid 35 m in diameter with a velocity of 5 km/s (Jewitt et al., 2011). The fragments smaller than 100 m in diameter and the dust cloud of micron-size particles with a mass of approximately $4 \times 10^7$ kg were formed as a result. After such a collision, a crater roughly 300 m in diameter could appear on the asteroid's surface (Bodewits et al., 2011). In the images of some asteroids directly taken from spacecraft, craters that are bluer than the surrounding surface are seen (Chapman, 1996). Most of the surface of asteroids is covered by regolith that became dark-red due to space-weathering processes, while blue regions on asteroids are associated with recent ejections of the internal fresher material (Chapman, 1996).

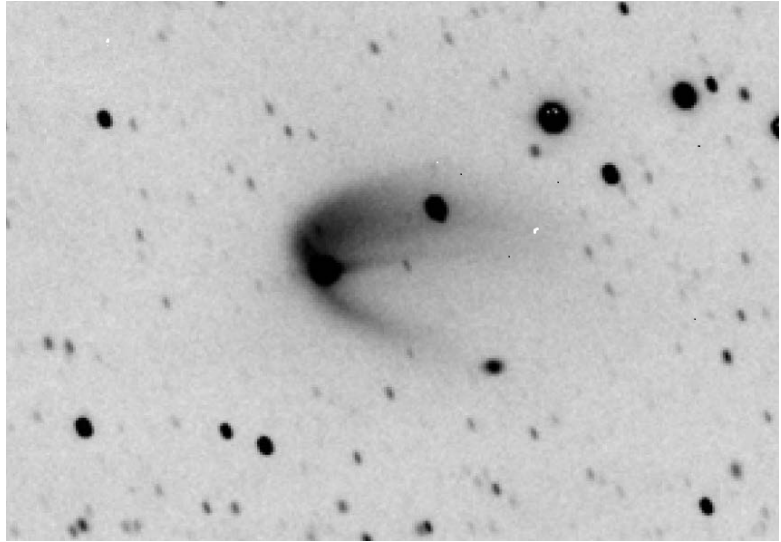

**Fig. 1.** The combined image of asteroid (596) Scheila during the outburst in December 2010 (A. Gibbs and S. Larson; credit https://uanews.arizona.edu).

From the ground-based observations carried out in December 2010 to May 2011, it was shown that the morphology of the asteroid's outburst changed for this time period; and the diameter and the mass of the ejected dust cloud were estimated. It was also supposed that the outburst could be caused by a collision with a large meteoroid that belongs to meteoroid streams related to two periodic comets (127P and P/2005K3), near which the asteroid flew by shortly before the outburst (Neslushan et al., 2016). Based on the observations performed in May–June 2011, Kiselev et al. (2014) showed that the asteroid continues to exhibit the same apparent brightness and, consequently, the collision with a small body in December 2010 did not result in substantial fragmentation of the object.

A purpose of new observations of asteroid Scheila is to continue the analysis of photometric characteristics of the asteroid—the brightness, the light curve shape, and the color indices—and to reveal their probable changes that could occur due to the collision, as well as to estimate the diameter of the object and its rotation period.

## OBSERVATIONS, PROCESSING, AND RESULTS

The astrometric and photometric observations of asteroid Scheila were carried out at the Zeiss-1000 telescope of the Sanglokh International Astronomical Observatory (SIAO) of the Institute of Astrophysics, Academy of Sciences, Republic of Tajikistan on June 16–17 and from July 30 to August 1, 2017. The telescope was equipped with a FLI Proline PL16803 CCD camera; the focal distance of the telescope (the Cassegrain focus) is $F = 13.3$ m, and the scale of the image is 63 μm/arcsec. The size and the field of view of the matrix are 4096 × 4096 pixels and $11' \times 11'$, respectively; the pixel scale of the matrix is $0.18''$ per pixel. To reduce the information redundancy and improve the signal-to-noise ratio $S/N$, the images were stored with using a binning value of 2, which made the working scale of the images to be $0.36''$ per pixel. The image quality, which was measured as an average of the full width at half maximum (FWHM) for several stars from individual images, was at a level of $2.1''$. We used the standard BVRI filters that rather closely reproduce the filters of the Johnson–Cousins photometric system. To reduce the noise level of the CCD matrix, the latter was cooled to a temperature of $-20°$C. To take into account the dark current, align the flat fields of the images, and take into account the CCD camera errors, we used the "Dark", "Flat", and "Bias" exposures, respectively; the latter were also used in the image processing. For the whole observational period, 550 images with exposure times of 10 to 60 s were taken. To make absolute photometric measurements, all of the field stars, which had been preliminarily examined for variability, were considered. The catalog of the Photometric All-Sky Survey of the American Association of Variable Star Observers (abbreviated as APASS; https://www.aavso.org/apass) was used for photometric studies. The APASS catalog contains the stars with magnitudes in the interval between $7^m$ and $17^m$ measured in five transmission bands: B and V of the Johnson–Cousins system and $g'$, $r'$, and $i'$ of the Sloan system. To pass from the Sloan system to the Johnson–Cousins one, the transition equations for the R band were taken from the paper by Mallama (2014). To process the images and calculate the sky background, we used the IDL codes

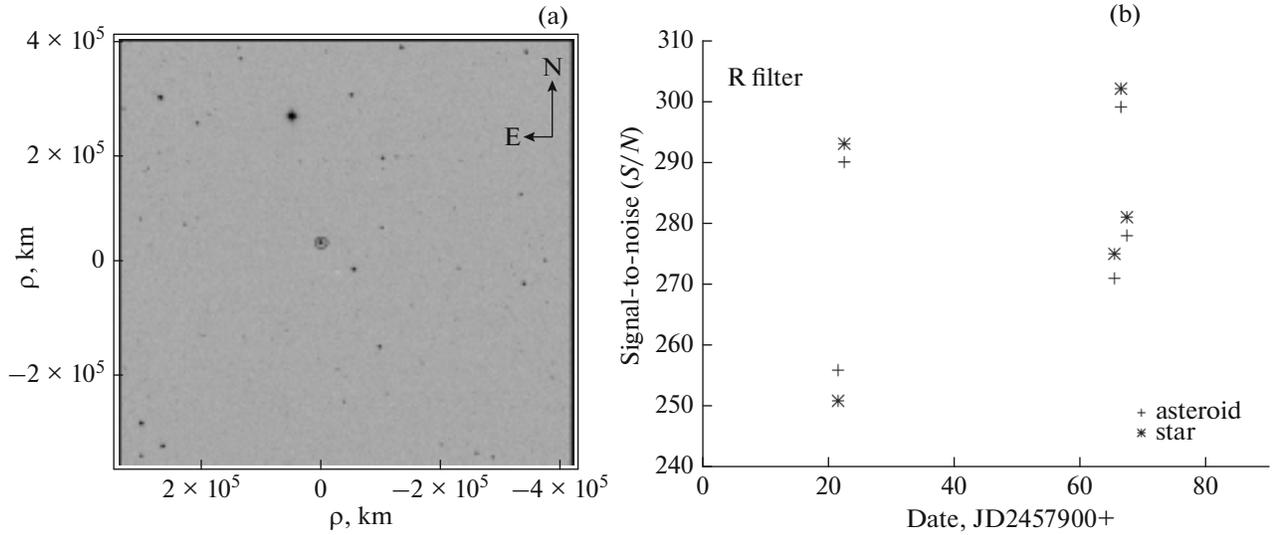

**Fig. 2.** The image of asteroid (596) Scheila in the R filter obtained at the Sanglokh observatory on June 16, 2017 (a) and the $S/N$ ratio for the asteroid and the standard star (for the images taken in the $R$ filter) (b).

(http://www.harrisgeospatial.com/SoftwareTechnology/IDL.aspx).

The image of asteroid Scheila is presented in Fig. 2a. To measure the images of the asteroid and the field stars, we used the fixed radius aperture, which allowed the object to be completely covered. In Fig. 2b, the $S/N$ values of the asteroid and the standard star are shown in dependence on the observational date for the images taken in the R filter. The dates and time of observations of the asteroid (expressed in the UT day fractions), the number of acquired images $N$, the exposure time $t$, as well as the geocentric $r$ and heliocentric $\Delta$ distances and the phase angle $ph$ of the asteroid are given in Table 2.

In terms of astrometry, the SIAO observations in the period from July 30 to August 1, 2017, were processed with the use of the APEKS-II software package developed at the Pulkovo Observatory (Devyatkin et al., 2010). The package calibrates the exposures, distinguishes the images of the stars and the objects, and identifies the stars according to the specified catalogs. In the astrometric reduction process, the image distortions produced by the optical system are taken into account with the six- and eight-constant methods (depending on a number of the identified stars). The astrometric UCAC4 catalog was used as a reference. The coordinates of stars with a magnitude between $10^m$ and $14^m$ are given in the UCAC4 with an astrometric accuracy of approximately $0.02''$, while the accuracy in the position of fainter stars (the limit is $16^m$) is roughly $0.07''$. To measure the positions in the frames, the reference stars with the brightness corresponding to the specified interval were used. The astrometric-reduction error averaged over all exposures is $0.323''$ and $0.190''$ for the right ascension $\alpha$ and the declination $\delta$, respectively. In Table 3, the mean deviations of the measured equatorial coordinates O from the catalog data $C$ designated as $(O-C)_\alpha$ and $(O-C)_\delta$ for the coordinates $\alpha$ and $\delta$, respectively, and their corresponding mean-square errors $\sigma_\alpha$ and $\sigma_\delta$ are presented according to the SIAO observations in the period from July 30 to August 1, 2017. The results of determining the coordinates of the asteroid according to the SIAO observations are shown in Fig. 3, where the right ascension $\alpha$ and the declination $\delta$ of the asteroid are plotted on the abscissa and the ordinate, respectively (http://www.minorplanetcenter.net/iau, M.P.C. 808122, 2017).

**Table 2.** Summary of observations of asteroid (596) Scheila at the Sanglokh observatory

| Data, UT | $r$, AU | $\Delta$, AU | $ph$, degree | $t$, s | N | | | |
|---|---|---|---|---|---|---|---|---|
| | | | | | B | V | R | I |
| June 16.85, 2017 | 2.452 | 1.473 | 8.2 | 10 | 43 | | 42 | |
| June 17.73, 2017 | 2.453 | 1.478 | 8.6 | 10 | | 100 | | |
| July 30.69, 2017 | 2.470 | 1.849 | 21.7 | 60 | 36 | 35 | 40 | 35 |
| July 31.71, 2017 | 2.471 | 1.861 | 21.8 | 60 | | 13 | | |
| August 1.65, 2017 | 2.471 | 1.873 | 21.9 | 60 | | 20 | | |

**Table 3.** The mean differences (O−C) and their standard deviations for asteroid (596) Scheila

| Date | (O−C)$_\alpha$, arcsec | $\sigma_\alpha$, arcsec | (O−C)$_\delta$, arcsec | $\sigma_\delta$, arcsec |
|---|---|---|---|---|
| July 30, 2017 | −0.038 | 0.026 | −0.016 | 0.043 |
| July 31, 2017 | −0.027 | 0.062 | −0.003 | 0.048 |
| Aug. 1, 2017 | −0.038 | 0.018 | −0.043 | 0.021 |

**Table 4.** Comparison of the initial orbit of asteroid (596) Scheila obtained from the Sanglokh observations to the MPC orbit (J2000.0)

| Orbital elements | SIAO (this study) | MPC | ε |
|---|---|---|---|
| Number of positions used for orbit calculations | 208 | 3758 | − |
| T | JD 2458716.1 | JD 2457892.4 | − |
| Epoch | 2457979.5 | 2458000.5 | − |
| e | 0.16354 | 0.16348 | −0.00006 |
| a, AU | 2.92821 | 2.92761 | −0.00060 |
| q, AU | 2.44932 | 2.44898 | −0.00034 |
| i, degree | 14.66023 | 14.66149 | 0.00126 |
| ω, degree | 70.60387 | 70.60631 | 0.00244 |
| Ω, degree | 175.27739 | 175.15669 | −0.12070 |
| n, degree/day | 0.19670 | 0.19676 | 0.00006 |
| σ | 0″.462 | 0″.356 | − |

From the coordinates of the asteroid measured in several positions, its orbit can be constructed. There are many methods for this, and they are used depending on the positions of the points measured on the orbit and the perturbations taken into account. First, from several observations, the initial orbit is constructed; and it is defined more precisely later, with the use of additional new observations. The orbit of Scheila was determined with the EPOS software package developed also at the Pulkovo Observatory (L'vov and Tsekmeister, 2012). For the mean time moment of 208 observations performed at the SIAO, we succeeded in obtaining the initial orbit of the asteroid. It is presented in Table 4. The following orbital elements are given for the 2000.0 equinox: the eccentricity $e$, the semimajor axis $a$, the perihelion distance $q$, the inclination $i$, the argument of perihelion $\omega$, the longitude of ascending node $\Omega$, the mean-diurnal motion $n$, and the mean-square errors $\varepsilon$. For comparison, in Table 4, we list the orbital elements calculated on the basis of the Minor Planet Center (MPC) database (http://www.minorplanetcenter.net/iau, M. P. C. 421631, 2017) from more than 3700 observations of the asteroid performed at different observatories from 1906 to 2017; the last column contains the corrections.

To process the photometric images in a standard way, we made master-frames of the zero exposure and dark and flat fields. All the frames containing the asteroid's images were corrected for zero-point and nonuniformity of the pixel sensitivity with the use of the master-frames. The sky background was determined with the standard IDL procedure called Sky (Landsman, 1993). For the aperture photometry of stars, the diaphragm with a radius of 7″ (3 × FWHM) was used. The residual sky background was estimated

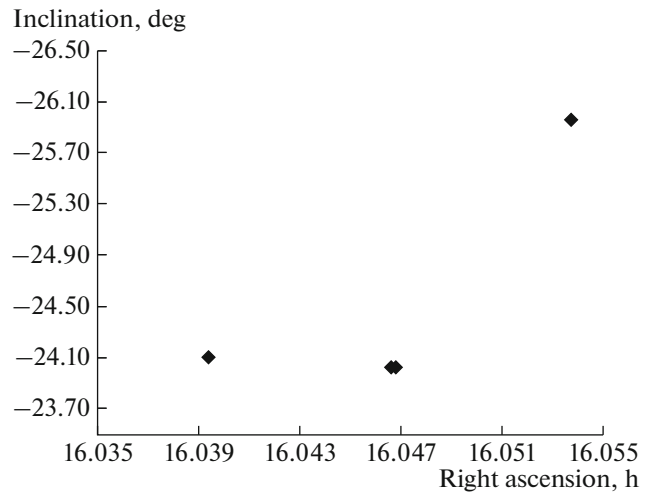

**Fig. 3.** The apparent trajectory of asteroid (596) Scheila according to the Sanglokh observations carried out in the period from July 30 to August 1, 2017.

**Table 5.** The apparent brightness (expressed in stellar magnitudes*) of asteroid (596) Scheila according to the Sanglokh observations in 2017

| Filter | June 16, 2017 | June 17, 2017 | July 30, 2017 | July 31, 2017 | Aug. 1, 2017 |
|---|---|---|---|---|---|
| B | 12.91 ± 0.04 | 12.92 ± 0.04 | 13.94 ± 0.04 | 13.98 ± 0.04 | 13.95 ± 0.04 |
| V | 12.24 ± 0.03 | 12.19 ± 0.03 | 13.22 ± 0.03 | 13.27 ± 0.03 | 13.24 ± 0.03 |
| R | 11.95 ± 0.03 | 11.91 ± 0.03 | 12.93 ± 0.03 | 12.96 ± 0.03 | 12.95 ± 0.03 |
| I | 11.63 ± 0.03 | 11.60 ± 0.03 | 12.61 ± 0.03 | 12.65 ± 0.03 | 12.63 ± 0.03 |

* The stellar magnitude was estimated within the aperture radius (the aperture size is 7″, which corresponds to 2692.21 km).

with a circular aperture. To calculate the stellar magnitude error, we summed up the statistical errors, which are caused by the S/N ratio for the object and the reference stars, and the errors in the catalog magnitudes of standard stars (for the APASS catalog, the error for the standard stars was assumed as $0.03^m$ (Henden et al., 2011)).

The light curves of asteroid Scheila obtained in such a way are shown in Fig. 4, where the apparent magnitudes $m$ and the observation dates (expressed in Julian days) are on the ordinate and the abscissa, respectively. The mean values of the apparent magnitudes of the object obtained in different filters are presented in Table 5.

As seen from Fig. 4 and Table 5, the brightness of the asteroid remains almost constant; its oscillations hardly exceed $0.1^m$. However, the rotation phases of the object covered by our monitoring are ignored here, which makes it difficult to draw a conclusion about the total amplitude of changes in the brightness of the asteroid. The visible brightness gradually decreased for the whole observational period, since the asteroid moved away from both the Sun and the observer. However, as will be shown below, the absolute brightness varied near the ephemeris value within the measurement errors.

The apparent magnitudes $m$ were converted to the absolute values $H$ according to the following semiempirical formula (Bowell et al., 1989) that allows the change in the asteroid's brightness to be described more precisely in the phase angle range from 0° to 120°.

$$H = m - 5\log(r\Delta) + 2.5\log[(1-G)\Phi_1 + G\Phi_2],$$
$$\Phi_i = \exp\left[-A_i\{\tan(\beta/2)\}^{B_i}\right], \quad i = 1, 2, \quad (1)$$

where $G$ is the slope parameter (the value of which was found only for several asteroids and assumed at 0.15 for the others), $\Phi_1$ and $\Phi_2$ are the functions of a phase angle, and $A_1 = 3.33$, $A_2 = 1.87$, $B_1 = 0.63$, and $B_2 = 1.22$ are the coefficients, the values of which are estimated by Penttila et al. (2016). The absolute brightness of the asteroid obtained in such a way in the $V$ and $R$ filters, $H_V$ and $H_R$ (the averages for one night), and the apparent brightness in these filters, $m_V$ and $m_R$, are listed in Table 6. The geocentric $r$ and heliocentric $\Delta$ distances and the phase angle $ph$ of the asteroid are also presented. The obtained absolute values vary from $(9.0 \pm 0.05)^m$ to $(8.8 \pm 0.03)^m$ in the $V$ and $R$ filters, respectively.

To determine the diameter of asteroid Scheila $D$ (expressed in kilometers), we used the following empirical expression, which was assumed for estimating asteroid sizes (Harris, 2002)

$$D = \frac{1329}{\sqrt{p_v} \times 10^{0.2H}}, \quad (2)$$

where $p_v = 0.038$ is the geometric albedo of asteroid Scheila (Tedesco and Desert, 2002). The estimates of the asteroid diameter according to the measurements of brightness $H$ in the $R$ filter are presented in Table 6 together with the other data. The effective diameter calculated with formula (2) from our measurements varies from 118.5 ± 1.6 to 124.1 ± 1.7 km, and these values are close to the available estimates based on different observations (including spaceborne ones) (see, Tedesco and Desert (2002) and references in Table 6).

The color index of the object, its albedo, and the characteristics of the spectrum of the reflected solar light are the important quantities, from which the taxonomic type of the asteroid and, consequently, its composition can be determined. On the basis of the low albedo value, $p_v = 0.038$ (Tedesco and Desert, 2002), the color index values $B-V = 0.71$ and $U-B = 0.18$ (http://ssd.jpl.nasa.gov, 2017), and the spectral characteristics, Scheila can be classified as an asteroid of the primitive P- or D-type; its mean volume density is 2 g/cm³, which is characteristic of carbonaceous material (Dahlgren and Lagerkvist, 1995). Note that, according to the measurements of the meteorite samples available, such density is also typical of carbonaceous chondrites (see, e.g., Consolmagno and Britt, 1998; Consolmagno et al., 2008).

The mean values of the color index of the object according to our and other observations are listed in Table 7. As is seen, the color indices for July 30, 2017, are rather close to those obtained in the observations on December 27, 2010, and January 4, 2011, which were carried out 16 and 24 days after the Scheila outburst. The color indices according to the observations performed almost right after the Scheila outburst on December 15, 2010, stand out from the others: these data do not correspond to any taxonomic class known (Betzler et al., 2012). The reported data show that the values of the color indices of Scheila were restored after the outburst rather rapidly, and their current values again correspond to those of the P- and D-type asteroids.

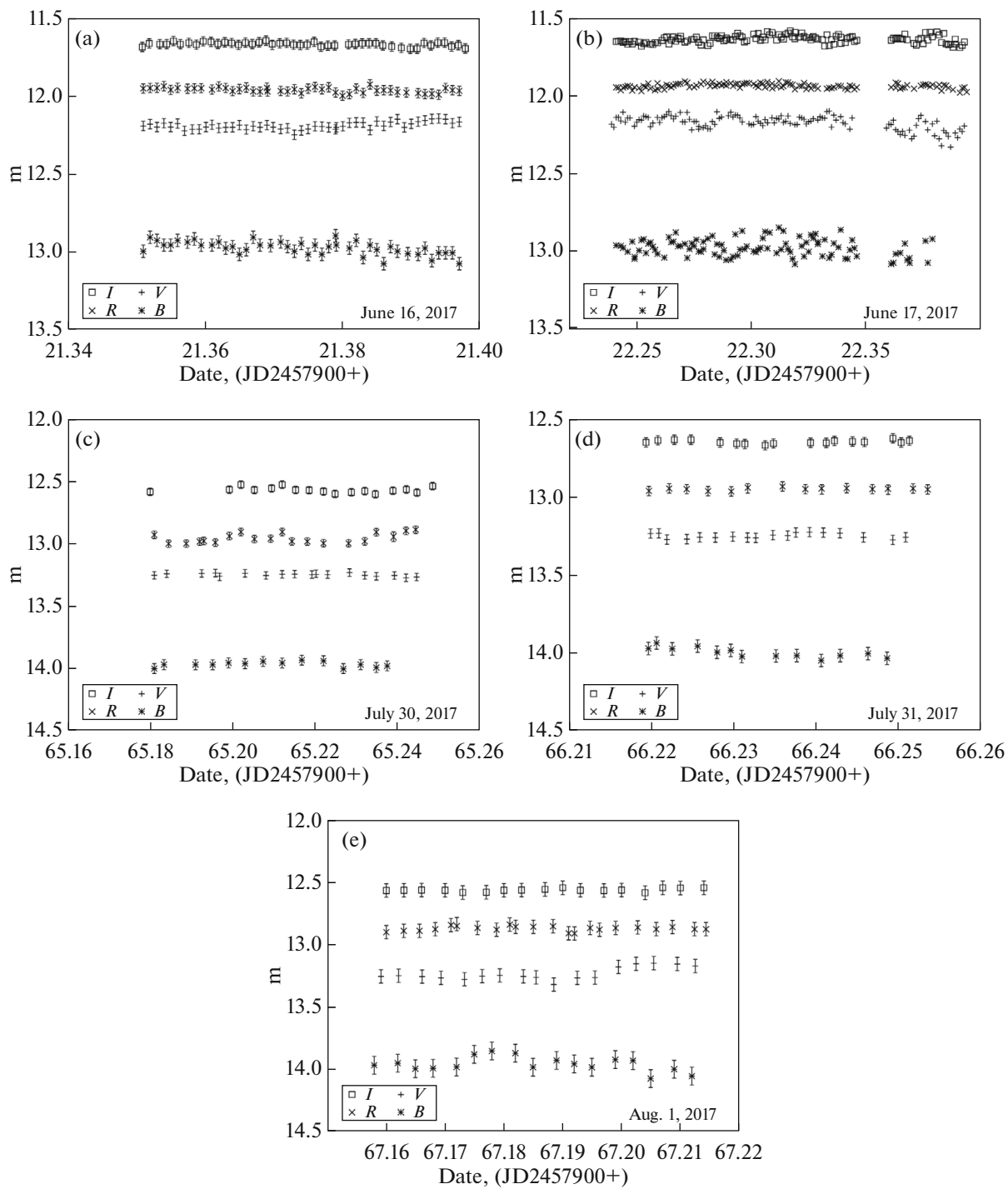

**Fig. 4.** The light curves of asteroid (596) Scheila in the BVRI filters according to the Sanglokh observations carried out on June 16–17 (panels (a) and (b)) and from July 31 to August 1, 2017 (panels c–e).

We also estimated the rotation period of the asteroid. To search for periodic changes in the brightness, we used the Period04 code based on the Fourier anal-ysis (https://www.univie.ac.at/tops/Period04) and all the data on the brightness in the I filter. This yielded the power spectrum (Fig. 5), where the $0.50^m$ amplitude corresponds to a rotation period of $16.1 \pm 0.2$ h; this estimate is close to 15.8 h, the value accepted for the rotation period (http://ssd.jpl.nasa.gov, 2017; http://www.minorplanetcenter.net/iau, 2017).

**Table 6.** The geometric aspect, the apparent and absolute brightness in the V and R filters, and the diameter of asteroid (596) Scheila according to the SIAO observations

| Date, UT | $r$, AU | $\Delta$, AU | $ph$, degree | $m_V$, magnitude | $m_R$, magnitude | $H_V$, magnitude | $H_R$, magnitude | $D$, km |
|---|---|---|---|---|---|---|---|---|
| June 16.85, 2017 | 2.452 | 1.473 | 8.2 | 12.24 ± 0.03 | 11.95 ± 0.03 | 9.1 ± 0.03 | 8.8 ± 0.03 | 118.5 ± 1.6 |
| June 17.73, 2017 | 2.453 | 1.478 | 8.6 | 12.19 ± 0.03 | 11.91 ± 0.03 | 9.0 ± 0.03 | 8.8 ± 0.03 | 118.5 ± 1.6 |
| July 30.69, 2017 | 2.470 | 1.849 | 21.7 | 13.22 ± 0.03 | 12.93 ± 0.03 | 9.1 ± 0.03 | 8.8 ± 0.03 | 118.5 ± 1.6 |
| July 31.71, 2017 | 2.471 | 1.861 | 21.8 | 13.27 ± 0.03 | 12.96 ± 0.03 | 9.1 ± 0.03 | 8.8 ± 0.03 | 118.5 ± 1.6 |
| Aug. 1.65, 2017 | 2.471 | 1.872 | 22.0 | 13.24 ± 0.03 | 12.95 ± 0.03 | 9.0 ± 0.03 | 8.7 ± 0.03 | 124.1 ± 1.7 |
| Jan.–Aug. 2010 | – | – | – | – | – | – | – | 159.7 ± 1.1 (Masiero et al., 2012) |
| Feb. 15.64, 2010 | 3.93 | 3.14 | 16.8 | – | 11.00 | – | 8.7 | 118 ± 6 (Bauer et al., 2012) |
| Dec. 27.90, 2010 | 3.085 | 2.338 | 13.7 | – | 13.98 | – | 8.85 | 113 ± 2 (Jewitt, 2012) |
| Jan. 4.90, 2011 | 3.073 | 2.254 | 11.9 | – | 13.86 | – | 8.86 | 113 ± 2 (Jewitt, 2012) |
| May 6.80, 2011 | 2.877 | 2.753 | 20.4 | – | 14.63 | – | 9.36 | 102 ± 3 (Neslushan et al., 2016) |
| May 8.80, 2011 | 2.873 | 2.776 | 20.4 | – | 14.43 | – | 9.40 | 102 ± 3 (Neslushan et al., 2016) |
| May 22.80, 2011 | 2.850 | 2.932 | 20.0 | – | 14.31 | – | 9.19 | 102 ± 3 (Neslushan et al., 2016) |
| Dec. 14.28, 2010 | – | – | – | | – | | | 118 ± 6 (Bodewits et al., 2011) |

**Table 7.** The mean color indices of asteroid (596) Scheila

| Date, UT | $B–V$ | $V–R$ | $R–I$ |
|---|---|---|---|
| June 16.85, 2017 | 0.71 ± 0.05 | 0.29 ± 0.03 | 0.32 ± 0.03 |
| June 17.73, 2017 | 0.73 ± 0.05 | 0.28 ± 0.03 | 0.31 ± 0.03 |
| July 30.69, 2017 | 0.72 ± 0.05 | 0.29 ± 0.03 | 0.32 ± 0.03 |
| July 31.71, 2017 | 0.71 ± 0.05 | 0.31 ± 0.03 | 0.31 ± 0.03 |
| Aug. 1.65, 2017 | 0.71 ± 0.05 | 0.29 ± 0.03 | 0.32 ± 0.03 |
| Dec. 27.90, 2010 (Jewitt, 2012) | 0.71 ± 0.03 | 0.38 ± 0.03 | – |
| Jan. 4.90, 2011 (Jewitt, 2012) | 0.71 ± 0.03 | 0.38 ± 0.03 | – |
| Dec. 15.49, 2010 (Neslushan et al., 2016) | 0.52 ± 0.004 | 0.49 ± 0.006 | 0.38 ± 0.009 |
| Sun | 0.64 ± 0.02 | 0.35 ± 0.01 | 0.33 ± 0.01 |

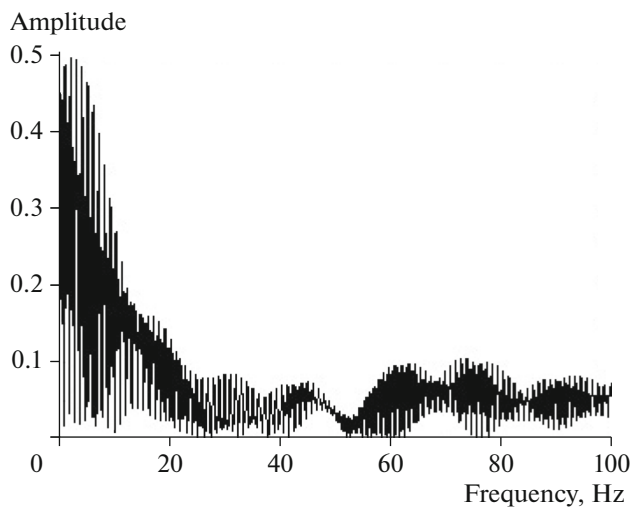

**Fig. 5.** The power spectrum based on the Fourier analysis of the Sanglokh observations of asteroid (596) Scheila in 2017.

## CONCLUSIONS

The astrometric and photometric observations of asteroid (596) Scheila carried out at the Sanglokh observatory in 2017 yielded the following results:

(1) The equatorial coordinates and the apparent trajectory of the asteroid were determined.

(2) During the observational period, the light curves showed no noticeable oscillations in the apparent brightness within the error measurements.

(3) The mean value of the absolute brightness of the asteroid in the V and R filters was $(9.1 \pm 0.05)^m$ and $(8.8 \pm 0.03)^m$, respectively.

(4) The diameter was estimated in the interval from $118.5 \pm 1.6$ to $124.1 \pm 1.7$ km.

(5) The color indices agree well with the currently available mean values for asteroids of the P- and D-types.

(6) The rotation period of the asteroid was estimated at $16.1 \pm 0.2$ h.

Thus, the asteroid continues to exhibit the earlier values of the absolute brightness and the other characteristics in spite of the collision with a small body undergone in December 2010. This allows us to suppose that the collision did not lead to catastrophic changes in the surface of the asteroid or to its substantial fragmentation. Most likely, the asteroid remained almost the same, while the consequences of the impact manifested themselves in cratering on the surface and ejecting a huge dust cloud that was evident as cometary activity.

## ACKNOWLEDGMENTS

O.V. Ivanova acknowledges the European Union's Seventh Framework Program (SASPRO) (grant no. 609427) and the Slovak Academy of Sciences (the Vega 2/2023/18 grant).